\documentclass[seceq,preprint]{ptptex}

\title{
A Generalized Gauge Theory of Gravity
}

\author{
Kouzou \textsc{Nishida}%
\footnote{E-mail: EZF01671@nifty.com} 
}

\inst{
Department of Physics, Kyoto Sangyo University, Kyoto 603-8555, Japan 
}

\abst{
In this paper, we discuss a gravitational theory based on the generalized
gauge field. 
Our Lagrangian is invariant not only under 
local Lorentz transformation and the ordinary gauge transformation 
but also under a new gauge transformation. We show that 
the gauge field associated with this new transformation is a second-rank 
tensor field and that the Einstein-Hilbert term can be derived from 
our Lagrangian when the gauge field has a vacuum expectation 
value. We also show that our model provides a Lagrangian for the scalar-tensor theory.
}

\begin{document}

\maketitle

\section{Introduction}
A considerable number of previous studies have explored the origins 
of scalar fields. In one of the previous studies, a model that regards 
a scalar field as a type of gauge field was considered. For example, 
the number of dimensions considered in the ``gauge-Higgs unification''\cite{rf:1,rf:2} was 
greater than four, and the extra components in a higher-dimensional 
gauge field were identified as the Higgs fields. 
In such a model, the scalar fields are constrained by the gauge principle, 
and thus, the predictive power of the model is at least as good as 
that of the Standard Model.

Sogami\cite{rf:4} proposed a covariant derivative of the form
\[
 \partial_{\mu}-\frac{1}{2}(1-\gamma_5)A_{L\mu}-\gamma_{\mu}\phi,
\]
where $\phi$ is the Higgs field. 
Note that the fields in this covariant derivative contain 
$\gamma_5$ and $\gamma_{\mu}$.
The gravitational field is a gauge field whose generators are 
a product of gamma matrices.\cite{rf:2-1} 
In our previous paper\cite{rf:3}, we considered the generalized 
gauge field that has various combinations of gamma matrices as generators 
and showed that the generalized 
gauge field includes vector field, spin connections, scalar field, 
and tensor field. Moreover, we identified the derived scalar as the Higgs 
field and attempted to extend the Standard Model. However, 
we were unable to determine the gauge transformation with 
which the gauge field of the scalar field or tensor field was associated. 
Moreover, our model was invariant under local Lorentz transformation; 
however, no discussion in terms of gravitational theory was 
provided in the paper. 

In this paper, we further generalize the gauge field discussed 
in the previous paper and attempt to apply it to gravitational theory 
in four and five dimensions. 
The results show that our model has a new gauge invariance,  
and the gauge field associated with the transformation is a second-rank tensor 
field. We show that if the tensor 
field has a vacuum expectation value, the Einstein-Hilbert term 
can be derived from our Lagrangian. 
We also show that our model can be used to explain 
the ``scalar-tensor theory.''\cite{rf:5,rf:6}

\section{The four-dimensional model}

We consider the following Lagrangian
\begin{equation}
 \label{eq:102}
  {\cal L}=\sqrt{-g}\left\{-\bar{\psi}i\gamma^{\mu}{\cal D}_{\mu}\psi
  -\frac{1}{4}{\rm Tr}( {\cal F}^{\mu\nu}{\cal F}_{\mu\nu}) \right\}
\end{equation}
with
\begin{equation}
 \label{eq:202}
  {\cal D}_{\mu} \equiv \partial_\mu-ie{\cal A}_{\mu}
  ,\,\,\,\,\,\,\,\,\,\,\,\,
  {\cal F}_{\mu\nu}\equiv\frac{i}{e}[{\cal D}_{\mu},{\cal D}_{\nu}],
\end{equation}
where Tr is the trace of a 4$\times$4 matrix and ${\cal A}_{\mu}$ 
is any 4$\times$4 matrix that satisfies 
\begin{equation}
 \label{eq:302}
  \gamma_{0}{\cal A}_{\mu}^{\dagger}\gamma_{0}=-{\cal A}_{\mu}.
\end{equation}
We adopt the metric signature $(-,+,+,+)$.

We find that (\ref{eq:102}) is invariant under the following transformations:
\begin{equation}
 \label{eq:402}  
  \psi \rightarrow \psi'=U(x)\psi 
  ,\,\,\,\,\,\,\,\,\,\,\,\,
  \gamma_{\mu} \rightarrow \gamma'_{\mu}=U(x)\gamma_{\mu} U^{-1}(x), 
\end{equation}
\begin{equation}
 \label{eq:502}  
 {\cal A}_\mu \rightarrow {\cal A}'_\mu=U(x){\cal A}_{\mu}U^{-1}(x)
 +\frac{1}{ie}(\partial_{\mu}U(x))U^{-1}(x),
\end{equation}
where $U(x)$ is an arbitrary matrix that satisfies
\begin{equation}
 \label{eq:602}
  \gamma_{0}U^{\dagger}(x)\gamma_{0}=-U^{-1}(x).
\end{equation}
Note that $\gamma'_{\mu}$ satisfies the relations,
$\{\gamma'_{\mu},\gamma'_{\nu} \}=2g_{\mu\nu}$, 
when 
$\{\gamma_{\mu},\gamma_{\nu} \}=2g_{\mu\nu}$.

The transformation $U(x)$ involves gauge transformation,
\begin{equation}
 \label{eq:702}
  U(x)=e^{iI\varepsilon(x)/2},
\end{equation}
and local Lorentz transformation,
\begin{equation}
 \label{eq:802}
  U(x)=e^{\gamma^{ij}\varepsilon_{ij}(x)/16}
  \,\,\,\,\,\,\,\,\,\,\,\,
  {\rm with}\,\,\gamma^{ij}\equiv \frac{\gamma^{i}\gamma^{j}-\gamma^{j}\gamma^{i}}{2},
\end{equation}
where $I$ denotes the $4\times 4$ unit matrix.
In addition, $U(x)$ involves the following new transformation:
\begin{equation}
 \label{eq:902}
  U(x)=e^{i\gamma^{i}\varepsilon_{i}(x)/2}.
\end{equation}
In general, $U(x)$ involves local transformations with the generators:
\begin{equation}
 \label{eq:1002}
  I,
  \,\,\,\,\,\,\,\,
  \gamma^i, 
  \,\,\,\,\,\,\,\,
  \gamma^{ij}, 
  \,\,\,\,\,\,\,\,
  \gamma_5\gamma^{i},
  \,\,\,\,\,\,\,\,
  \gamma_5.
\end{equation}
Here, we have constructed a flat tangent space at every point on 
the four-dimensional manifold, 
and 
we indicate the vectors in the five-dimensional tangent space 
by subscript Roman letters $i$, $j$, $k$, etc. 
Greek letters like $\mu$ and $\nu$ are used to
label four-dimensional space-time vectors.
Note that $\gamma_i$ is a constant matrix.
Now, we define the {\it vierbein} (tetrad for four-legged) 
that is used for the transformation 
from the $x$-space to the tangent space, 
and vice versa, as
\begin{eqnarray}
 \label{eq:1102}
  & &b^{i}\,_{\mu}b_{i\nu}=g_{\mu\nu},
  \,\,\,\,\,\,\,\,\,\,\,\,
  b^{i\mu}=g^{\mu\nu}b^{i}\,_{\nu},
  \nonumber\\
  & &b^{i}\,_{\mu}b_{j}\,^{\mu}=\delta^{i}_{j}.
\end{eqnarray}
We can then define a set of gamma matrices over either the
tangent space or the $x$-space as
\begin{equation}
 \label{eq:1202}
  \gamma^{i}b_i\,^{\mu}=\gamma^{\mu},
  \,\,\,\,\,\,\,\,\,\,\,\, 
  \{\gamma^\mu,\gamma^\nu \}=2g^{\mu\nu}.
\end{equation}

Because (\ref{eq:102}) is invariant under gauge transformation and 
local Lorentz transformation, ${\cal A}_{\mu}$ 
must involve the gauge field and the spin connection. 
It is known that an arbitrary $4\times4$ matrix $M$ can be decomposed
into 16 gamma matrices, that is,
$I$, $\gamma^i$, 
$\gamma^{ij}$, 
$\gamma_5\gamma^{i}$, and $\gamma_5$, as
\begin{eqnarray}
 \label{eq:1302}
  M
  &=&
  \frac{1}{4}\left[{\rm Tr}(M)I+{\rm Tr}(\gamma_{5}M)\gamma_5
  +{\rm Tr}(\gamma_{i}M)\gamma^{i}-{\rm Tr}(\gamma_5\gamma_iM)\gamma_5\gamma^i\right]\nonumber\\
  & &
  -\frac{1}{8}{\rm Tr}(\gamma_{ij}M)\gamma^{ij}.
\end{eqnarray}
By substituting $M={\cal A}_{\mu}$ in (\ref{eq:1302}), we derive
\begin{eqnarray}
 \label{eq:1402}
  {\cal A}_{\mu}
  &=&\frac{1}{2}\left( A_{\mu}I+A^{(5)}_{\mu}\gamma_{5}+A_{i\mu}\gamma^{i}
  -A^{(5)}_{i\mu}\gamma_{5}\gamma^{i}+\frac{i}{2}\omega_{ij,\mu}\gamma^{ij} \right).
\end{eqnarray}
Here, we define
\begin{eqnarray}
 \label{eq:1502}
  \lefteqn{ A_{\mu}\equiv \frac{1}{2}{\rm Tr}({\cal A}_{\mu}),
  \,\,\,\,\,\,\,\,\,\,\,\,
  A^{(5)}_{\mu}\equiv \frac{1}{2}{\rm Tr}(\gamma_{5}{\cal A}_{\mu}), }
  \nonumber\\
  & &A_{i\mu}\equiv \frac{1}{2}{\rm Tr}(\gamma_{i}{\cal A}_{\mu}),
  \,\,\,\,\,\,\,\,\,\,\,\,
    A^{(5)}_{i\mu}\equiv \frac{1}{2}{\rm Tr}(\gamma_{5}\gamma_{i}{\cal
    A}_{\mu}),
  \nonumber\\
  & &\omega_{ij,\mu}\equiv \frac{i}{2}{\rm Tr}(\gamma_{ij}{\cal A}_{\mu}).
\end{eqnarray}

Let us investigate how each field in (\ref{eq:1502}) changes under the
infinitesimal transformation described in 
(\ref{eq:702}), (\ref{eq:802}), and (\ref{eq:902}).\\
For (\ref{eq:702}), we consider the trace of ${\cal A}'_{\mu}$ in
(\ref{eq:502}) and obtain
\begin{equation}
 \label{eq:1602}
  \delta A_{\mu}=\frac{1}{e}\partial_{\mu}\varepsilon.
\end{equation}
This implies that $A_{\mu}$ is the usual gauge field. \\
For (\ref{eq:802}), we multiply ${\cal A}'_{\mu}$ in (\ref{eq:502}) 
from the left by $\frac{i}{2}\gamma_{ij}$ 
and consider the trace to obtain
\begin{equation}
 \label{eq:1702}
  \delta
  \omega_{ij,\mu}=\varepsilon_{i}\,^{k}\omega_{kj,\mu}+\varepsilon_{j}\,^{k}\omega_{ik,\mu}
  -\frac{1}{4e}\partial_{\mu}\varepsilon_{ij}.
\end{equation}
From the abovementioned equation, 
we can identify $\omega_{ij,\mu}$ as the spin connection. \\
For (\ref{eq:902}), we have
\[
 \delta A_{i\mu}=\varepsilon^{k}\omega_{ik,\mu}+\frac{1}{e}\partial_{\mu}\varepsilon_{i},
 \,\,\,\,\,\,\,\,\,\,\,\,
 \delta \omega_{ij,\mu}=\varepsilon_{i}A_{j\mu}-\varepsilon_{j}A_{i\mu},
\]
\begin{equation}
 \label{eq:1802}
  \delta A^{(5)}_{i\mu}=i\varepsilon_{i}A^{(5)}_{\mu},
  \,\,\,\,\,\,\,\,\,\,\,\,
  \delta A^{(5)}_{\mu}=i\varepsilon^{k}A^{(5)}_{k\mu}.
\end{equation}
In other words, $A_{i\mu}$ is a gauge field that is associated with the new
transformation described in (\ref{eq:902}).

All the fields represented by (\ref{eq:1502}) couple with 
a universal coupling constant $e$. 
Hence, each field represented by (\ref{eq:1502}) 
is interpreted as a type of gravitational field.

We can reexpress ${\cal L}$ also in terms of (\ref{eq:1402}).
The calculation, however, can be markedly simplified by extending our model to five dimensions.
In the next section, we will show the extension of the present model to five dimensions 
and discuss the features of our model in detail.

\section{The five-dimensional model}

We consider the following action
\begin{equation}
 \label{eq:103}
  S=\int d^{5}x\sqrt{-g}\left\{-\bar{\psi}i\gamma^{M}{\cal D}_{M}\psi
  -\frac{\chi}{4}{\rm Tr}( {\cal F}^{MN}{\cal F}_{MN}) \right\}
\end{equation}
with
\begin{equation}
 \label{eq:203}
  {\cal D}_{M} \equiv \partial_M-ie{\cal A}_{M}
  ,\,\,\,\,\,\,\,\,\,\,\,\,
  {\cal F}_{MN}\equiv\frac{i}{e}[{\cal D}_{M},{\cal D}_{N}],
\end{equation}
where $M=0,1,2,3,4$ and ${\cal A}_{M}$ 
is any 4$\times$4 matrix that satisfies 
\begin{equation}
 \label{eq:303}
  \gamma_{0}{\cal A}_{M}^{\dagger}\gamma_{0}=-{\cal A}_{M},
\end{equation}
and $\gamma^{M}$ represents the five-dimensional gamma matrices,
\begin{equation}
 \label{eq:403}
  \gamma^{M}=(\gamma^{\mu},\gamma^{4})
  \,\,\,\,\,\,\,\,\,\,\,\,
  {\rm with}\,\,\gamma^{4}\equiv \gamma_{5}.
\end{equation}
Since ${\cal A}_{M}$ has a mass dimension of 1, as seen from (\ref{eq:203}), 
we multiply the bosonic part of the Lagrangian by a coefficient $\chi$ 
whose mass dimension is 1, so that $S$ becomes dimensionless.

(\ref{eq:103}) is invariant under the following transformations:
\begin{equation}
 \label{eq:503}  
  \psi \rightarrow \psi'=U\psi 
  ,\,\,\,\,\,\,\,\,\,\,\,\,
  \gamma_{M} \rightarrow \gamma'_{M}=U\gamma_{M} U^{-1}, 
\end{equation}
\begin{equation}
 \label{eq:603}  
  {\cal A}_M \rightarrow {\cal A}'_M
  =U{\cal A}_{M}U^{-1}+\frac{1}{ie}(\partial_{M}U)U^{-1},
\end{equation}
where
\begin{equation}
 \label{eq:703}
  U=e^{iI\varepsilon(x)/2},
  \,\,\,\,\,
  e^{\gamma^{ab}\varepsilon_{ab}(x)/16},
  \,\,\,\,\,
  e^{i\gamma^{a}\varepsilon_{a}(x)/2}.
\end{equation}
Here, we indicate the vectors in the five-dimensional tangent space 
by subscript Roman letters $a$, $b$, $c$, etc. 
Moreover, we use capital letters such as $M$ and $N$ 
to label the five-dimensional space-time vectors.

Equation (\ref{eq:1302}) can be rewritten as
\begin{equation}
 \label{eq:803}
  M
  =
  \frac{1}{4}\left[{\rm Tr}(M)I
  +{\rm Tr}(\gamma_{a}M)\gamma^{a}\right]
  -\frac{1}{8}{\rm Tr}(\gamma_{ab}M)\gamma^{ab}.
\end{equation}
By substituting $M={\cal A}_{M}$ in (\ref{eq:803}), we obtain 
\begin{equation}
 \label{eq:903}
  {\cal A}_{M}=\frac{1}{2}(A_{M}I+A_{aM}\gamma^{a}+\frac{i}{2}\omega_{ab,M}\gamma^{ab}).
\end{equation}
Here, we define
\begin{equation}
 \label{eq:1003}
  A_{M}\equiv \frac{1}{2}{\rm Tr}({\cal A}_{M}),
  \,\,\,\,\,\,\,\,\,\,\,\,
  A_{aM}\equiv \frac{1}{2}{\rm Tr}(\gamma_{a}{\cal A}_{M}),
  \,\,\,\,\,\,\,\,\,\,\,\,
  \omega_{ab,M}\equiv \frac{i}{2}{\rm Tr}(\gamma_{ab}{\cal A}_{M}).
\end{equation}
Note that $A^{(5)}_{\mu}$ and $A^{(5)}_{i\mu}$ in the four-dimensional
model have been included in $A_{aM}$ and $\omega_{ab,M}$, respectively.

Using (\ref{eq:903}), we can calculate the field strength as
\begin{eqnarray}
 \label{eq:1103}
  {\cal F}_{MN} &=& \frac{i}{e}[{\cal D}_{M},{\cal D}_{N}] \nonumber \\
  &=& \frac{1}{2}(D_{M}A_{aN}-D_{N}A_{aM})\gamma^{a}
  -\frac{i}{2}eA_{aM}A_{bN}\gamma^{ab}
  \nonumber \\
  & &+\frac{i}{4}R_{ab,MN}\gamma^{ab}+\frac{1}{2}F_{MN}I,
\end{eqnarray}
where
\[
 D_{M}A_{aN} \equiv \partial_{M}A_{aN}
 +e\omega_{a}\,^{b}\,_{,M}A_{bN},
\]
\[
 R^{ab}\,_{,MN} \equiv 
 \partial_{M}\omega^{ab}\,_{,N}-\partial_{N}\omega^{ab}\,_{,M} 
 +e(\omega^{a}\,_{c,M}\omega^{cb}\,\,_{,N}-\omega^{a}\,_{c,N}\omega^{cb}\,\,_{,M}) ,
\]
\begin{equation}
 \label{eq:1203}
  F_{MN}\equiv \partial_{M}A_{N}-\partial_{N}A_{M}.
\end{equation}
We know that the Riemann curvature tensor $R^{P}\,_{Q,MN}$ can be written as
\begin{equation}
 \label{eq:1303}
  R^{P}\,_{Q,MN} = b_{a}\,^{P}b_{bQ}R^{ab}\,\,_{,MN}.
\end{equation}

Using this field strength, we determine the bosonic part of the Lagrangian:
\begin{eqnarray}
 \label{eq:1403}
  \lefteqn{
  -\frac{1}{4}{\rm Tr}({\cal F}^{MN}{\cal F}_{MN}) 
  }
  \nonumber \\
  &=&
  \frac{1}{4}eA_{a}\,^{M}A_{b}\,^{N}R^{ab}\,_{,MN}
  -\frac{1}{4}(D_{M}A_{aN}-D_{N}A_{aM})(D^{M}A^{aN}-D^{N}A^{aM})
  \nonumber \\
  & &
  -\frac{1}{4}e^{2}(A_{aM}A^{aM}A_{bN}A^{bN}-A_{aM}A^{aN}A_{bN}A^{bM})
  \nonumber \\
  & &-\frac{1}{8}R^{CD}\,_{,MN}R_{CD,}\,^{MN}
  -\frac{1}{4}F^{MN}F_{MN}.
\end{eqnarray}

We intend to include the Ricci curvature $R$ in (\ref{eq:1403}).
Now, we propose two methods to obtain $R$.
The first method involves the use of an expectation value.
Let us assume that $ A_{aM}$ has an expectation value given by
\begin{equation}
 \label{eq:1313}
  \langle A_{aM} \rangle=\frac{1}{2}v(y) b_{aM}
\end{equation}
and write 
\begin{equation}
 \label{eq:1323}
  A_{aM}=\frac{1}{2}v(y) b_{aM}+A'_{aM},
\end{equation}
where $v(y)$ is a function of the fifth coordinate $y$.
The validity of the vacuum expectation value 
will be discussed in the last section.
In general, the covariant derivative of the {\it funfbein} 
(tetrad for five-legged) should be zero,
\begin{equation}
 \label{eq:1603}
  \partial_{M}b_{aN}+e\omega_{ab,M}b^{b}\,_{N}-\Gamma^{P}\,_{NM}b_{aP}=0,
\end{equation}
where $\Gamma^{C}\,_{NM}$ is the Christoffel symbol. 
Using (\ref{eq:1323}) and (\ref{eq:1603}), we obtain 
\begin{eqnarray}
 \label{eq:1713}
  \lefteqn{
  -\bar{\psi}i\gamma^{M}{\cal D}_{M}\psi
  }
 \nonumber \\
 & = &
  -\bar{\psi}i\gamma^{M}\partial_{M}\psi-\frac{5}{4}ev(y)\bar{\psi}\psi
  -\frac{e}{2}A'_{M}\,^{M}\bar{\psi}\psi
  +\frac{e}{2}A'_{MN}\bar{\psi}\gamma^{MN}\psi
  +\cdots,
\end{eqnarray}
and
\begin{eqnarray}
 \label{eq:1813}
  \lefteqn{
  -\frac{1}{4}{\rm Tr}({\cal F}^{MN}{\cal F}_{MN}) 
  }\nonumber \\
  &=&
  \frac{1}{16}ev^{2}(y)R
  -\frac{1}{2}\partial^{4}v(y)\partial_{4}v(y)
  -\frac{1}{16}v^{2}(y)C^{P}\,_{,MN}C_{P,}\,^{MN}
  -\frac{1}{4}v(y)(\partial^{4}v(y))C^{M}\,_{,4M}
  \nonumber \\
  & &
  -\frac{5}{16}e^{2}v^{4}(y)
  -\frac{1}{16}e^{2}v^{2}(y)(6A'_{MN}A'^{MN}
  +4A'_{M}\,^{M}A'_{N}\,^{N}-2A'_{MN}A'^{NM})
  \nonumber \\
  & &-\frac{1}{8}R^{PQ}\,_{,MN}R_{PQ,}\,^{MN}
  -\frac{1}{4}F^{MN}F_{MN}
  +\frac{1}{4}eA'_{P}\,^{M}A'_{Q}\,^{N}R^{PQ}\,_{,MN}+\cdots,
\end{eqnarray}
where $C^{P}\,_{,MN}$ is the torsion:
\begin{equation}
 \label{eq:1903}
  C^{P}\,_{,MN} \equiv
  \Gamma^{P}\,_{MN}-\Gamma^{P}\,_{NM}.
\end{equation}
Note that we obtain the Ricci curvature $R=b_{a}\,^{M}b_{b}\,^{N}R^{ab}\,_{,MN}$.

When $v(y)$ is constant,
if we choose a value of $e$ such that 
\begin{equation}
 \label{eq:2203}
  \frac{1}{16}ev^{2}=\frac{1}{16\pi G},
\end{equation}
we obtain the Einstein-Hilbert term. From (\ref{eq:1713}), 
the fermion mass $m_{F}$ can be defined as
\begin{equation}
 \label{eq:2303}
  m_{F}^{2}=\frac{25}{16}e^{2}v^{2}=\frac{25}{2}eM_{p}^{2},
\end{equation}
where 
$M_{p}^{2} \equiv \frac{1}{8\pi G}$ is the Planck mass.
Since $M_{p}^{2}$ is very large, 
the coupling $e$ must be very small to determine 
the fermion mass, which is in the order of GeV.

Let us assume that $v(y)$ takes the following form,
\begin{equation}
 \label{eq:2313}
  v(y)=v_{0}\sin \frac{2\pi}{a}y,
\end{equation}
where $v_{0}$ is a constant 
with a mass dimension of 1, 
and $a$ is the radius of the fifth dimension.
When the fifth dimension is curled into a very small circle,
the mass of the zero-mode fermion is zero
owing to $\int^{a}_{0}\sin \frac{2\pi}{a}ydy=0$.
In this case, the condition (\ref{eq:2303}) is unnecessary.

Equation (\ref{eq:1813}) has a quadratic term 
$R^{NC}\,\,_{,MN}R_{CD,}\,\,^{MN}$; 
this quadratic term is not observed in the Einstein theory.
However, it is known that when the quantum effect is considered, 
higher order curvature tensors are observed in the effective Lagrangian.\cite{rf:8}

The second method for obtaining $R$ is based on conformal transformation.
First, let us decompose $A_{aM}$ as
\begin{equation}
 \label{eq:2313}
  A_{aM}=\frac{1}{2}\phi b_{aM}+A'_{aM},
\end{equation}
where
$\phi$ is a real scalar field with a mass dimension of 1 and 
\begin{equation}
  A'_{aM} \equiv A_{aM}-\frac{1}{2}\phi b_{aM}.
\end{equation}
By substituting (\ref{eq:2313}) into (\ref{eq:103}) 
and using (\ref{eq:1603}), we obtain
\begin{eqnarray}
 \label{eq:1703}
  \lefteqn{
  -\bar{\psi}i\gamma^{M}{\cal D}_{M}\psi
  }
\nonumber \\
 & = &
  -\bar{\psi}i\gamma^{M}\partial_{M}\psi-\frac{5}{4}e\phi\bar{\psi}\psi
  -\frac{e}{2}A'_{M}\,^{M}\bar{\psi}\psi
  +\frac{e}{2}A'_{MN}\bar{\psi}\gamma^{MN}\psi
  +\cdots,
\end{eqnarray}
and
\begin{eqnarray}
 \label{eq:1803}
  \lefteqn{
  -\frac{1}{4}{\rm Tr}({\cal F}^{MN}{\cal F}_{MN}) 
  }\nonumber \\
  &=&
  \frac{1}{16}e\phi^{2}R
  -\frac{1}{2}\partial^{M}\phi\partial_{M}\phi
  -\frac{1}{16}\phi^{2}C^{P}\,_{,MN}C_{P,}\,^{MN}
  -\frac{1}{4}\phi(\partial^{M}\phi)C^{N}\,_{,MN}
  -\frac{5}{16}e^{2}\phi^{4}
  \nonumber \\
  & &-\frac{1}{8}R^{PQ}\,_{,MN}R_{PQ,}\,^{MN}
  -\frac{1}{4}F^{MN}F_{MN}
  + \frac{1}{4}eA'_{P}\,^{M}A'_{Q}\,^{N}R^{PQ}\,_{,MN}+\cdots.
\end{eqnarray}
The first term on the right-hand side of (\ref{eq:1803}) is 
called nonminimal coupling. 
In other words, we obtain a Lagrangian of the scalar-tensor theory. 
We can show that a conformal transformation
\begin{equation}
 g_{MN}  \rightarrow g_{*MN}=\Omega^{2}(x)g_{MN}
\end{equation}
gives
\begin{equation}
 R=\Omega^{2}(x)(R_{*}+\cdots),
  \,\,\,\,\,\,\,\,\,\,\,\,
 \sqrt{-g}=\Omega^{-5}(x)\sqrt{-g_{*}},
\end{equation}
where $\Omega(x)$ is an arbitrary function 
of the space-time coordinate $x$. 
From Reference 9), if we choose $\Omega(x)$ such that
\begin{equation}
 \frac{1}{16}e\phi^{2}\Omega^{-3}(x)=\frac{1}{16\pi G}
\end{equation}
and introduce a scalar field $\sigma(x)$ that satisfies
\begin{equation}
 \phi=\frac{2\sqrt{2}}{\sqrt{e}}M_{p}e^{\zeta\sigma}
  \,\,\,\,\,\,\,\,\,\,\,\,
  {\rm with}
  \,\,\,\,\,\,\,\,\,\,\,\,
 \zeta^{-2} \equiv \left( \frac{16}{3}+\frac{8}{e} \right) M^{2}_{p}, 
\end{equation}
we can obtain 
\begin{equation}
 \sqrt{-g}\left( \frac{1}{16}e\phi^{2}R
 -\frac{1}{2}\partial^{M}\phi\partial_{M}\phi \right)
 =\sqrt{-g_{*}}\left( \frac{1}{16\pi G}R_{*}
 -\frac{1}{2}\partial^{M}\sigma\partial_{M}\sigma \right),
\end{equation}
and
\begin{equation}
 \label{eq:2323}
  \sqrt{-g}\frac{5}{4}e\phi\bar{\psi}\psi
  =\sqrt{-g_{*}}\frac{5\sqrt{2e}}{2}M_{p}e^{\zeta\sigma}\bar{\psi_{*}}\psi_{*},
\end{equation}
where we apply the transformation,
\begin{equation}
 \psi \rightarrow \psi_{*}=\Omega^{2}(x)\psi,
  \,\,\,\,\,\,\,\,\,\,\,\,
 \bar{\psi}\rightarrow \bar{\psi_{*}}=\Omega^{2}(x)\bar{\psi}
\end{equation}
to $\psi$ and $\bar{\psi}$.
Note that the Einstein-Hilbert term and the mass term 
for the fermion are obtained simultaneously.

Reference 9) or 10) shows that the interaction term including the scalar
field in Jordan frame can be eliminated from the Lagrangian in Einstein
frame, if the Lagrangian possesses scale invariance in Jordan frame.
The complete decoupling of the scalar field from the matter 
in Einstein frame does imply the weak equivalence principle (WEP),
whereas (\ref{eq:2323}) includes the coupling. 
Hence, our model might potentially violate WEP.
This issue will remain a subject of future studies.

\section{Discussion}

Here, we discuss the validity of the vacuum expectation value 
in (\ref{eq:1313}).
Equation (\ref{eq:1813}) includes a quartic term of $v(y)$. 
However, for making an assumption that the local minimum of the potential is not zero, 
the quadratic terms of $v(y)$ must also be considered. 
The quadratic terms may be induced by radiative correction,\cite{rf:9,rf:10} 
or the relevant $v^2(y)$ terms in (\ref{eq:1813}),
\begin{equation}
 \label{eq:104}
  \frac{1}{16}ev^{2}(y)R
  -\frac{1}{2}\partial^{4}v(y)\partial_{4}v(y)
  -\frac{1}{16}v^{2}(y)C^{C}\,_{,MN}C_{C,}\,^{MN}
  -\frac{1}{4}v(y)(\partial^{4}v(y))C^{N}\,_{,4N}
\end{equation}
may induce the same effect as the quadratic terms.\cite{rf:11}

For example, let us extend our model to six dimensions. 
If the extra space is a two-dimensional curved space 
of constant positive curvature $a^{-2}$, 
the six-dimensional Ricci curvature $R$ can be written as
\begin{equation}
 \label{eq:204}
  R=R_4+\frac{1}{a^2},
\end{equation}
where $R_4$ represents the four-dimensional Ricci curvature.
Then, we have the following quadratic term:
\begin{eqnarray}
 \label{eq:304}
  {\cal L}_{6}&=&\frac{1}{16}ev^{2}(y)R-\frac{5}{16}e^{2}v^{4}(y)\cdots
  \nonumber \\
  &=&
  \frac{1}{16}ev^{2}(y)R_{4}+\frac{1}{16a^2}ev^{2}(y)-\frac{5}{16}e^{2}v^{4}(y)
  \cdots.
\end{eqnarray}
Note that the relative sign of the $v^{2}(y)$ and $v^{4}(y)$ terms
is negative. 
When $a$ assumes an extremely small value, $v(y)$ takes an enormous vacuum 
expectation value. 
Hence, it is possible 
that $A_{aM}$ takes the vacuum expectation value as (\ref{eq:1313}).

%


\begin{thebibliography}{99}


\bibitem{rf:1}
N.~S.~Manton, \NPB{158,1979,141}.

\bibitem{rf:2}
D.~B.~Fairlie, \PLB{82,1979,97}.

\bibitem{rf:4}
I.~S.~Sogami, \PTP{94,1995,117}.

\bibitem{rf:2-1}
R.~Utiyama, \PR{101,1956,1597}. 

\bibitem{rf:3}
K.~Nishida, \PTP{118,2007,903}.

\bibitem{rf:5}
P.~Jordan,\ {\it Schwerkraft\ und\ Weltall}\ 
(Friedrich\ Vieweg\ und\ Sohn,\ Braunschweig,\ 1955).

\bibitem{rf:6}
C.~Brans and R.~H.~Dicke, \PR{124,1961,925}. 

\bibitem{rf:8}
N.~D.~Birrell and P.~C.~W.~Davies,\ {\it Quantum\ Fields\ in\ Curved\ Space}\ 
(Cambridge\ University\ Press,\ Cambridge,\ 1982).

\bibitem{rf:7}
Y.~Fujii and K.~Maeda,\ {\it The\ Scalar-Tensor\ Theory\ of\ Gravitation}\ 
(Cambridge\ University\ Press,\ Cambridge,\ 2003).

\bibitem{rf:7-1}
Y.~Fujii, \PTP{118,2007,983}.

\bibitem{rf:9}
Y.~Hosotani, \PLB{126,1983,309}; Ann.\ of\ Phys.\ \textbf{190}\ (1989),\ 223. 

\bibitem{rf:10}
I.~Antoniadis, K.~Benakli and M.~Quiros, New\ J.\ Phys.\ \textbf{3}\ (2001),\ 20, hep-th/0108005.

\bibitem{rf:11}
T.~Hattori, M.~Hayashi, T.~Inagaki and Y.~Kitadono, hep-ph/0408220.


\end{thebibliography}
\end{document}